\documentclass[twocolumn,preprintnumbers,amsmath,amssymb,superscriptaddress]{revtex4}
\usepackage{array}
\usepackage{booktabs}
\usepackage{tabu}
\usepackage{dcolumn}
\usepackage{amsmath}
\usepackage{amsfonts}
\usepackage{amssymb}
\usepackage{graphicx}
\usepackage[colorlinks,linkcolor=red,citecolor=blue]{hyperref}
\usepackage{subfigure}
\usepackage{graphicx}% Include figure files
\usepackage{dcolumn}% Align table columns on decimal point
\usepackage{bm}% bold math
\newtheorem{thm}{Theorem}

\newcommand{\commentold}[1]{}
\DeclareMathSymbol{:}{\mathpunct}{operators}{"3A}

\def\be{\begin{equation}}
\def\ee{\end{equation}}
\def\bea{\begin{eqnarray}}
\def\eea{\end{eqnarray}}
\def\f{\frac}
\def\n{\nonumber}
\def\l{\label}

\begin{document}
%\preprint{APS/123-QED}

\title{Information and the second law of thermodynamics}
\author{B. Ahmadi}
\email{borhanahmadi19@yahoo.com}
\address{Department of Physics, University of Kurdistan, P.O.Box 66177-15175, Sanandaj, Iran}
\address{International Centre for Theory of Quantum Technologies, University of Gdansk, Wita Stwosza 63, 80-308 Gdansk, Poland}
\author{S. Salimi}
\email{shsalimi@uok.ac.ir}
\address{Department of Physics, University of Kurdistan, P.O.Box 66177-15175, Sanandaj, Iran}
\author{ A. S. Khorashad}
\address{Department of Physics, University of Kurdistan, P.O.Box 66177-15175, Sanandaj, Iran}

\date{\today}% It is always \today, today,

\def\br{\biggr}
\def\bl{\biggl}
\def\Br{\Biggr}
\def\Bl{\Biggl}
\def\be\begin{equation}
\def\ee{\end{equation}}
\def\bea{\begin{eqnarray}}
\def\eea{\end{eqnarray}}
\def\f{\frac}
\def\n{\nonumber}
\def\l{\label}
\begin{abstract}
The second law of classical thermodynamics, based on the positivity of the entropy production, only holds for deterministic processes. Therefore the Second Law in stochastic quantum thermodynamics may not hold. By making a fundamental connection between thermodynamics and information theory we will introduce a new way of defining the Second Law which holds for both deterministic classical and stochastic quantum thermodynamics. Our work incorporates information well into the Second Law and also provides a thermodynamic operational meaning for negative and positive entropy production.
\end{abstract}

%Use showkeys class option if keyword
\maketitle
\newpage
%======================================================================
%============== Introduction ==================================
%======================================================================
\section{Introduction}
Thermodynamics and information have intricate inter-relations. Soon after establishing the second law of thermodynamics by Rodulf Clausius, Lord Kelvin and Max Planck \cite{Kondepudi,Blundell,Callen}, in his 1867 thought experiment, ''Maxwell's Demon'', James Clerk Maxwell attempted to show that thermodynamics is not strictly reducible to mechanics \cite{Maxwell,Leff,Maruyama}. Although Maxwell introduced his demon to question the Second Law, established by others, his demon revealed the relationship between thermodynamics and information theory for the first time. Clausius \textit{et. al} never considered information playing any role in constituting the Second Law. Maxwell illustrated that by using information about the positions and momenta of the particles restrictions imposed by the Second Law can be relaxed thus demanding to take into account information in the Second Law explicitly. In order to do this we must elucidate the physical nature of information so that the Second Law includes information as a physical entity. In 1929 L\'{e}o Szil\'{a}rd \cite{Zeitschrift}, inspired by Maxwell's idea, designed an engine working in a cycle, interacting with a single thermal reservoir, which used information (gained by the measurement on the system) to perform work.
\newline
In classical thermodynamics the Clausius' statement of the Second Law implies that in an irreversible process the entropy production of a system is always positive which means that information is always lost or encoded and never regained or decoded. The second law of classical thermodynamics only holds for \textit{deterministic macroscopic} equilibrium systems \cite{Kondepudi,Blundell,Callen} and is not necessarily valid in \textit{non-equilibrium stochastic microscopic} systems \cite{Evans,Searles,Wang,An}. By equilibrium we mean that both the initial and final states of the system should be equilibrium thermal states. Therefore it is expected that the generalization of the second law of classical thermodynamics to stochastic quantum thermodynamics may expose features not present in a classical setting. It will be seen that the stochastic nature of quantum thermodynamics leads to the results different than that of classical ones. Our aim in this work is to introduce a new way of defining the Second Law which holds for both deterministic classical and stochastic quantum thermodynamics. In order to do this we make a connection between thermodynamics and information theory in a fundamental way. We will incorporate information into the Second Law, and then show that this relation is fundamental. Based on this relation a generic form for the efficiency of an engine working in an arbitrary cycle will be derived and we will clarify why and how backflow of information can cause a quantum engine to be more efficient than a Carnot engine. Our results provide a thermodynamic operational meaning for negative entropy production, which until now only had information-theoretical interpretations; for example, it witnesses the non-Markovianity (or backflow of information) \cite{Chen}. It will also be shown that a quantum thermodynamic force \cite{Ahmadi} decodes (encodes) information (not) to be used by the system to perform more work than what is expected and consequently the efficiency of the system exceeds the Carnot efficiency.
\section{Classical engines and its limitation}
\begin{figure}[h]
\centering
\includegraphics[width=4cm]{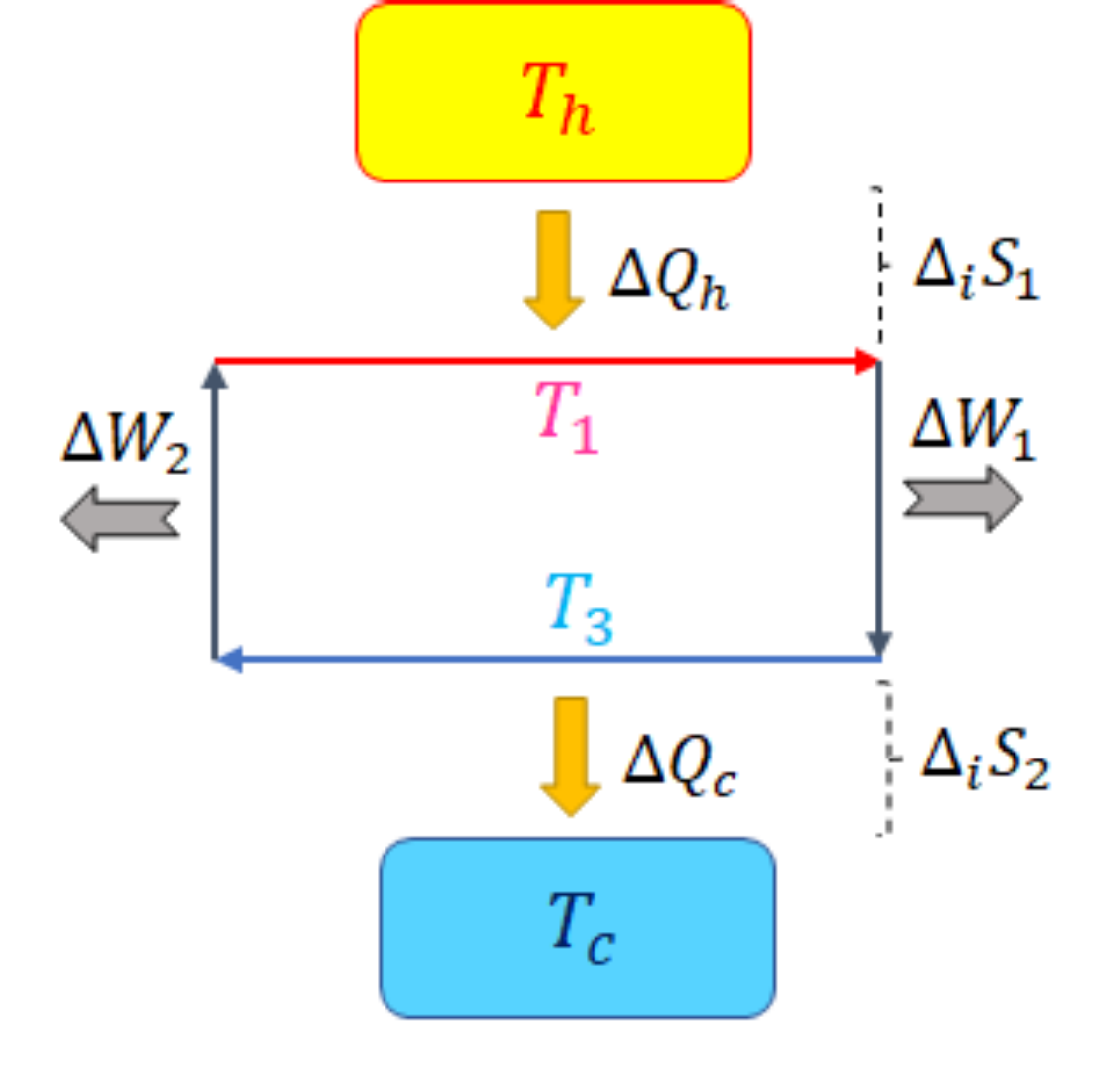}% Here is how to import EPS art
\caption{(Color online) A classical engine working between two reservoirs at temperatures $T_h>T_c$. Irreversibility occurs between the engine and the reservoirs not in the interior of the engine.}
\label{fig:Fig1.pdf}
\end{figure}

Before proceeding with the results let us examine a classical engine (see Fig. (\ref{fig:Fig1.pdf})) which gives us the motivation of having a quantum engine more efficient than Carnot classical engine. For this engine we obtain (see Supplementary Note \ref{Note1})
\begin{equation}\label{a0}
\eta_e-\eta_C=-\dfrac{T_c\Delta_iS}{\Delta Q_h},
\end{equation}
where $\eta_e=1-T_3/T_1$ is the engine efficiency, $\eta_C=1-T_c/T_h$ the Carnot efficiency and $\Delta_iS=\Delta_iS_1+\Delta_iS_2$ the entropy production of the total system (the engine plus the reservoirs) during a cycle. Based on the second law of classical thermodynamics the entropy production is always positive thus $\eta_e$ can never exceed $\eta_C$. This is, in fact, equivalent to the relation $T_c\leq T_3\leq T_1\leq T_h$ which always holds for classical engines. Now the question is: does this relation also always hold for quantum engines? If it does not, therefore quantum engines may be more efficient than that of Carnot. In the following we will investigate this question and see how it can be violated in the quantum realm hence demanding a new form of the Second Law.
\section{Reversible and Irreversible work}
The work done by a thermodynamic system, in the weak coupling limit, can always be appropriately partitioned into two parts: reversible work and irreversible work (see Supplementary Note \ref{Note2}),
\begin{equation}\label{eq:a1}
\Delta W=\Delta W_{rev}+\Delta W_{irr},
\end{equation}
in which
\begin{equation}\label{eq:a2}
\Delta W_{rev}\equiv\dfrac{1}{\beta}\Delta I+\Delta F^\beta,
\end{equation}
and
\begin{equation}\label{eq:a3}
\Delta W_{irr}\equiv\dfrac{1}{\beta}\Delta_iS.
\end{equation}
It is seen that the total work can always be partitioned into two parts, the reversible part and the irreversible part. This partitioning seems plausible since whenever the process is reversible all the work is reversible and there exists no irreversible work as expected. The reversible work defined in Eq. (\ref{eq:a2}) is different from the definition of reversible work in the literature, i.e., $\Delta W_{rev}\equiv\Delta F^\beta$ by the term $\dfrac{1}{\beta}\Delta I$. Classical thermodynamics is the equilibrium thermodynamics and the minimal work can be extracted only in equilibrium processes, hence the minimal work equals the equilibrium work, i.e., $\Delta W_{min}=\Delta F^\beta$. The term $\Delta I$ in Eq. (\ref{eq:a2}) explains the fact that quantum thermodynamics is a non-equilibrium thermodynamics, i.e., distance from the equilibrium state affects the minimal work that can be extracted. Another interesting point is that, as in equilibrium thermodynamics, based on our partitioning the amount of the internal energy which is encoded not to be used by the system to do work equals $\dfrac{1}{\beta}\Delta_iS$. This is notable because Eq. (\ref{eq:a3}) indicates the fact that regardless of the equilibrium or non-equilibrium thermodynamics the amount of the internal energy which is encoded equals $\dfrac{1}{\beta}\Delta_iS$. In other words the relation $\Delta W_{irr}=\dfrac{1}{\beta}\Delta_iS$ as a link between thermodynamics and information theory is fundamental. This relation will serve as an important building block in the rest of this investigation. We must note that $\Delta W_{rev}$ and $\Delta W_{irr}$ in Eq. (\ref{eq:a1}) are not done by the system during the process. The work which is done by the system is $\Delta W$. $|\Delta W_{rev}|$ is the maximal amount of the internal energy which is supposed to be spent by the system as work if there was no irreversibility during the process and positive $\Delta W_{irr}$ is the amount of the internal energy which is not allowed to be spent by the system as work due to irreversibility (loss of information). This explains why, regardless of the equilibrium or non-equilibrium thermodynamics, Eq. (\ref{eq:a3}) has the same form in both stochastic and deterministic thermodynamics and thus is fundamental. Hence the encoded internal energy which is not supposed to be used as work is always directly related to the entropy production (or information). In other words, we can say
\begin{equation}\label{eq:11aa}
\Delta_iS=\beta\times(internal\ enery\ not\ to\ be\ spent\ as\ work).
\end{equation}
We can go further and define a non-equilibrium free energy for a generic statistical state $\rho$ of a quantum system in contact with a thermal bath as
\begin{equation}\label{eq:11a}
F(\rho, H)\equiv E-TS=tr\{\rho H\}-TS(\rho),
\end{equation}
and we find
\begin{equation}\label{eq:11bb}
\dfrac{1}{\beta}\Delta_iS=\Delta W_{irr}=\Delta W-\Delta F.
\end{equation}
The associated non-equilibrium free energy is analogous to its equilibrium counterpart in non-equilibrium processes. As can be seen from Eq. (\ref{eq:11bb}) the minimal work, on average, necessary to drive the system from one arbitrary state to another is the difference, $\Delta F$, between the non-equilibrium free energy in each state. The excess work with respect to this minimum is the dissipated or irreversible work, $\Delta W_{irr}$. If the entropy production is positive $\Delta_iS\geq0$ then the generalized minimal work formulation (the generalized second law) for an isothermal process with given initial and final non-equilibrium distributions is obtained as
\begin{equation}\label{eq:11c}
\Delta W\geq\Delta F^\beta+\dfrac{1}{\beta}\Delta I.
\end{equation}
The generalized minimal work formulation of thermodynamics for non-equilibrium distributions gives an important relation between two major concepts in physics, energy and information. In the following we will show that in non-equilibrium quantum thermodynamics the internal energy can also be decoded (negative irreversible work) to be used by the system to perform more work than what is typically expected.
\section{Irreversible work and the Second Law}
\begin{figure}[h]
\centering
\includegraphics[width=4cm]{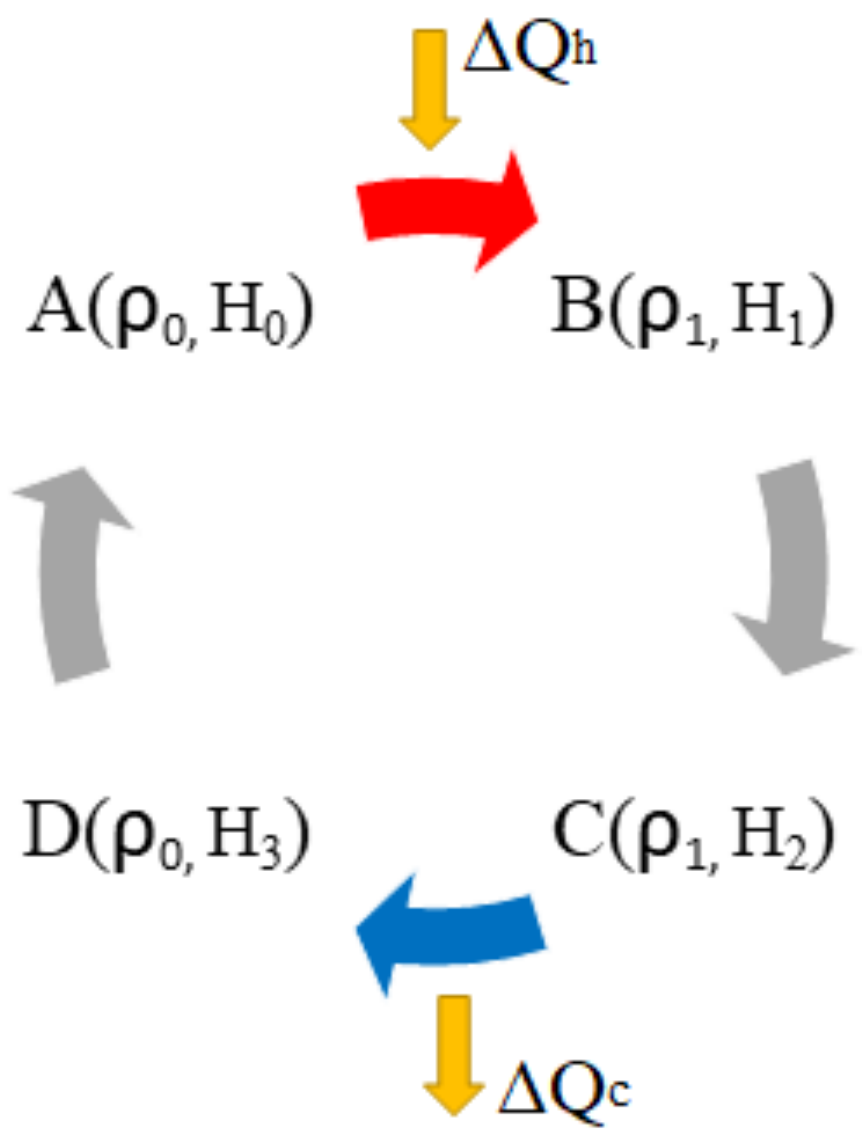}% Here is how to import EPS art
\caption{(Color online) As a visual aid points, between which the quantum system operates as the working substance in an Otto cycle, in dynamical configuration space are depicted in a $(\rho, H)$-coordinate system. From $A$ to $B$ the heat $\Delta Q_h$ is absorbed from the hot reservoir at temperature $T_h$ and from $C$ to $D$ the heat $\Delta Q_c$ is rejected to the cold reservoir at temperature $T_c$. The processes from $B$ to $C$ and from $D$ to $A$ occur adiabatically.}
\label{fig:Fig2.pdf}
\end{figure}

Let us consider a system in state $\rho_0$ at time $t=0$ attached to a bath of temperature $T$. After a finite-time $\tau$, let the state of the system be $\rho_\tau$. The Hamiltonian $H$ of the system remains unchanged during the evolution. The irreversible work after a time $\tau$ is obtained as (see Supplementary Note \ref{Note3})
\begin{equation}\label{eq:11d}
\Delta W_{irr}=\dfrac{1}{\beta}S(\rho_0\|\rho^\beta)-S(\rho_\tau\|\rho^\beta).
\end{equation}
During a Markovian evolution $\Delta W_{irr}$ is always positive but for a non-Markovian evolution it can be negative and this may lead to results not encountered in classical thermodynamics. In the following we focus our attention on four special cases to elucidate the physical meaning of the relation $\Delta W_{irr}=\dfrac{1}{\beta}\Delta_iS$ in non-equilibrium quantum thermodynamics:
 \newline
(a) Consider a reversible cycle with a quantum engine operating between two heat reservoirs at temperatures $T_h$ and $T_c$ with $T_h>T_c$. Since all the processes are reversible then the work done by the system is $\Delta W=\Delta W_{rev}$. For a machine to work as an engine we should have $\Delta W_{rev}<0$ and since the cycle is reversible, $\Delta W_{rev}= T_h\Delta S_h+T_c\Delta S_c$, the efficiency of the engine equals the Carnot efficiency,
\begin{equation}\label{eq:12}
\eta\equiv\dfrac{-\Delta W}{\Delta Q_h}= 1-\dfrac{T_c}{T_h}=\eta_C,
\end{equation}
where $\Delta Q_h$ is the heat absorbed from the hot reservoir. Eq. (\ref{eq:12}) holds for all reversible cycles with classical or quantum heat engines \cite{Quan}. In equilibrium thermodynamics the Clausius' statement of the Second Law leads to the fact that of all the heat engines working between two given temperatures, none is more efficient than a Carnot engine \cite{Kondepudi,Blundell,Callen}. As can be seen from Eq. (\ref{eq:a2}) the reason behind this is that in equilibrium thermodynamics, due to the Clausius' statement of the Second Law, the entropy production $\Delta_iS$ can never be negative, thus it can never help $-\Delta W$ to increase, i.e., the production of entropy is an indication of a reduction in the thermal efficiency of the engine. In the language of information the Clausius' statement of the Second Law means that information can never be decoded in deterministic thermodynamics. As we will show below, in stochastic quantum thermodynamics, this is also true as long as the process is Markovian. But if the process is non-Markovian $\Delta_iS$ can be negative \cite{Marcantoni,Chen}, i.e., some of the internal energy can be decoded to be used by the system as work and consequently the efficiency can exceed the Carnot efficiency.
\newline
(b) Consider a quantum engine operating in a cycle between two heat reservoirs at temperatures $T_h$ and $T_c$ with $T_h>T_c$. In step I, as depicted in Fig. (\ref{fig:Fig2.pdf}), the engine interacts with a hot reservoir at temperature $T_h$ from point $A(\rho_0, H_0)$ to point $B(\rho_1, H_0)$ while the Hamiltonian remains unchanged. The heat absorbed by the engine is $\Delta Q_h=tr\{H_0(\rho_1-\rho_0)\}$. In step II the engine is decoupled from the hot reservoir and undergoes an adiabatic evolution from point $B(\rho_1, H_0)$ to point $C(\rho_1, H_1)$. In step III it interacts with a cold reservoir at temperature $T_c$ from point $C(\rho_1, H_1)$ to point $D(\rho_0, H_1)$ while the Hamiltonian remains unchanged. The heat rejected to the cold reservoir is $\Delta Q_c=tr\{H_1(\rho_0-\rho_1)\}$. Finally in step IV the engine is decoupled from the cold reservoir and, in an adiabatic evolution, goes back to its initial point by going from point $D(\rho_0, H_1)$ to point $A(\rho_0, H_0)$ and complete the cycle. Now the whole work done by the system during the cycle, as in Eq. (\ref{eq:a2}), is $\Delta W=\Delta W_{irr}+\Delta W_{rev}$. Since during the adiabatic processes no entropy is produced in the interior of the system \cite{Gemmer,Quan} thus
\begin{equation}\label{eq:13}
\Delta W_{irr}=\dfrac{1}{\beta_h}\Delta_iS_h+\dfrac{1}{\beta_c}\Delta_iS_c.
\end{equation}
Then the efficiency becomes
\begin{equation}\label{eq:14}
\eta= \dfrac{-\Delta W_{rev}}{\Delta Q_h}-\dfrac{\dfrac{\Delta_iS_h}{\beta_h}+\dfrac{\Delta_iS_c}{\beta_c}}{\Delta Q_h}
%&=&\eta_C-\dfrac{\dfrac{\Delta_iS_h}{\beta_h}+\dfrac{\Delta_iS_c}{\beta_c}}{\Delta Q_h}.
\end{equation}
Eq. (\ref{eq:14}) is a generic form for the efficiency of any engine working in a cycle. In the case of a reversible cycle the second term on the right hand side vanishes and it simply reduces to Eq. (\ref{eq:12}). Now as is clear from Eq. (\ref{eq:14}) the second term on the right hand side shows the contributions of the Markovianity and non-Markovianity of the processes to the efficiency. If the evolution of the system during steps I and III is Markovian then $\Delta_iS_h$ and $\Delta_iS_c$ are positive and consequently decrease the efficiency which would be less than the Carnot efficiency. In the language of information this means that some information is encoded (lost) hence the system cannot use this encoded energy as work during the evolution. But if the evolution of the system during steps I and III is non-Markovian then $\Delta_iS_h$ and $\Delta_iS_c$ can be negative and consequently can increase the efficiency which can become greater than that of Carnot. In the language of information this means that the information is \textit{decoded} and the system uses this decoded information to perform additional work and as a result the efficiency increases. Thus, as was mentioned before, one way to exceed the Carnot efficiency is to have non-Markovian processes during the cycle. As an example, consider a spin-1/2 system \cite{Quan,Thomas,Thomas1,Kieu} working in an Otto cycle, as depicted in Fig. (\ref{fig:Fig2.pdf}). The system is in an initial state $\rho_0$, diagonal in the eigenbasis of the Hamiltonian $H_0=(\omega_0/2)\sigma_z$, where $\omega_0=\kappa B$ and $\sigma_z$ is the Pauli matrix. Here $\kappa$ is a constant and $B$ is the constant magnetic field applied in the $z$ direction on the system. The efficiency of the engine reads (see Supplementary Note \ref{Note4})
\begin{equation}\label{14a}
\eta=1-\dfrac{\omega_1}{\omega_0}\leq1-\dfrac{T_3}{T_1}.
\end{equation}
For a Markovian process in order to absorb heat from the hot reservoir and reject heat to the cold reservoir we must have $T_c\leq T_3\leq T_1\leq T_h$. Thus $1-T_3/T_1\leq1-T_c/T_h$, i.e., the efficiency is less than Carnot efficiency, and the Second Law is preserved. But in the case of non-Markovian baths, since the effective temperature of the system may not approach the temperature of the reservoir monotonically \cite{Xiong,Zhang,Paavola,Intravaia,Goan}, we can have $T_3\leq T_c\leq T_h\leq T_1$ which can lead to an engine more efficient than that of Carnot, resulting in the violation of the Second Law. It should be noted that decoupling the system from the reservoir might cause some energy cost. But since, in the example above, we are in the weak coupling limit and turning on and off the interaction occurs very fast compared to the time of the step, this energy cost is negligible, i.e, $E_{int}=tr\{\rho H_{int}\}\simeq0$.
\newline
It is worth mentioning that, although Clausius \textit{et. al} made no mention of information in establishing the Second Law, from our point of view the information had already been incorporated into the Second Law. Because positive entropy production means that information is encoded. In Ref. \cite{Deffner1} a device interacting with two heat reservoirs, a work reservoir, and an information reservoir which exchanges information but not energy with the device was investigated. They have found that for cyclic processes in which information is systematically written to the memory, the efficiency can exceed the Carnot limit. It should be noted that in this case the system and the bath are not left to themselves, i.e., the information reservoir acts as Maxwell's demon which intervene in the process from the outside to decode information. But in the case of non-Markovian bath nothing intervenes in the process from the outside, i.e., information is decoded without any help from the outside of the system and the bath.
\newline
(c) Consider a quantum engine interacting with only one heat reservoir at temperature $T$. The work done after a cycle is $\Delta W=-\Delta Q$ thus, using Eq. (\ref{eq:11g}) of supplementary Note \ref{Note3}, we get
\begin{eqnarray}\label{eq:14ab}\nonumber
\Delta W&=&\dfrac{1}{\beta}\Delta_iS_h+\dfrac{1}{\beta}\Delta_iS_c\nonumber \\
&=&T[S(\rho_1\|\rho_0^\beta)-S(\rho_0\|\rho_0^\beta)+S(\rho_1\|\rho_1^\beta)-S(\rho_0\|\rho_1^\beta)],\nonumber \\
\end{eqnarray}
where $\rho_0^\beta=\exp(-\beta H_0)/tr\{\exp(-\beta H_0)\}$ and $\rho_1^\beta=\exp(-\beta H_1)/tr\{\exp(-\beta H_1)\}$. For this cycle if the heat reservoir is Markovian Eq. (\ref{eq:14ab}) is positive, as we mentioned before, i.e., no negative work can be extracted which is in complete agreement with the Kelvin-Planck statement of the Second Law which asserts that no process is possible whose sole result is the extraction of energy from a heat bath, and the conversion of all that energy into work \cite{Kondepudi,Blundell,Callen}. But if the heat reservoir is non-Markovian Eq. (\ref{eq:14ab}) can be negative. In other words negative work can be extracted in a cycle with only one heat reservoir and this is strictly against the Second Law. This violation occurs just because the second law of deterministic classical thermodynamics is only based on losing information (flow of information) while in stochastic quantum thermodynamics information could also backflow into the system.
\newline
(d) As the last case let us now consider a quantum engine working in a cycle, similar to the one in case (a), between two heat reservoirs at temperatures $T_h$ and $T_c$, but with $T_h<T_c$, i.e., the engine transports heat from a cool reservoir to a hot reservoir and performs negative work. According to the second law of classical thermodynamics this is impossible and strictly violates the Clausius' statement of the Second Law which declares that ''No process is possible whose sole result is the transfer of heat from a colder to a hotter body'' \cite{Kondepudi,Blundell,Callen}. In this case $\Delta W_{irr}=\dfrac{1}{\beta_h}\Delta_iS_h+\dfrac{1}{\beta_c}\Delta_iS_c$ and if non-Markovianity is strong enough such that $\mid\Delta W_{irr}\mid>\mid\Delta W_{rev}\mid$ negative work can be output thus the efficiency of the engine could be greater than zero. Non-Markovianity to be strong enough means that enough information to be decoded to extract enough negative work. In Ref. \cite{Zhang} a quantum Otto heat engine has been investigated. They have found that if the heat reservoirs are Markovian $T_h$ must be larger than $2T_c$ in order for negative work to be output; however, in the non-Markovian case negative work can be performed if $T_h>0.8T_c$. From our point of view the condition $T_h>0.8T_c$ is the condition under which enough information is decoded for the system to perform negative work in the cycle.
\newline
The four cases considered above help to understand the physical nature of information. Rolf Landauer declared in 1991 that ''information is physical'' \cite{Landauer}. Since then, information has come to be seen by many physicists as a fundamental component of the physical world \cite{Karnani,Vedral,Vedral1,Davies}. In deterministic equilibrium thermodynamics we could also have negative entropy production. Szil\'{a}rd showed that information can be used to do work if one permits an intelligent being (demon) to intervene in the process of a thermodynamic system \cite{Zeitschrift}. In the language of information what Maxwell's demon does is that it decodes (gathers or brings back) information and the system uses this decoded information to output more work. Decoding information causes the entropy production of the system to be negative, therefore as we have shown above this causes the system to perform more work and, in turn, it leads to have an efficiency greater than that of Carnot. Del Rio \textit{et. al} \cite{Rio} have shown that erasing a system, which is coupled strongly with another system (a quantum memory), may cause the conditional entropy of the system to be negative and this negative entropy will lead to extracting work from the system, thus cooling the environment. Our results provide a thermodynamic operational meaning for negative entropy production, which until now only had information-theoretical interpretations; for example, it witnesses the non-Markovianity (or backflow of information). The significance of a general Szil\'{a}rd engine is that it conjoins thermodynamics and information theory. It shows the usefulness of information for performing some thermodynamic task. Given the important link between the task of work extraction and information theory, as appears in the examples of Maxwell's demon \cite{Sagawa}, the Szil\'{a}rd engine \cite{Zeitschrift}, and Landauer's erasure principle \cite{Landauer1}, it is becoming more common to consider the nature of information as physical. It is now well understood that the role of the demon does not contradict the second law of thermodynamics, because the initialization of the demon's memory entails heat dissipation \cite{Parrondo,Landauer1,Lloyd}. In classical thermodynamics if we leave the system to itself (i.e. no demon is allowed to intervene) there is no way to have a negative entropy production thus the Carnot engine is in fact the most efficient engine possible. But in the quantum realm, due to the existence of landmark quantum features, even if the system is left to itself, in non-Markovian processes the entropy production of the system can be negative (information is decoded) thus the system, working in a cycle, can be more efficient than a Carnot engine. Now we are in a position to properly define the second law of thermodynamics for both classical and quantum thermodynamics as:
\newline
\section{The second law of thermodynamics}
In a thermodynamic process information can be encoded and also decoded for the system to perform work which equals temperature $T$ times the entropy production of the system, i.e.,
\begin{equation}\label{eq:14b}\nonumber
\Delta W_{irr}=\dfrac{1}{\beta}\Delta_iS,
\end{equation}
where $\beta=1/T$ is the temperature of the reservoir with which the system interacts. As can be seen this definition of the Second Law emphasizes on the connection between thermodynamics (work as a thermodynamic variable) and information theory, not on a specific direction for the arrow of time because unlike deterministic classical thermodynamics in stochastic quantum thermodynamics the entropy production can be both positive and negative. This way of defining the Second Law covers both classical and quantum thermodynamics and also incorporates information well into the Second Law, i.e., it is never violated in the quantum realm nor in the presence of a demon intervening in the process. Therefore Carnot's, Clausius' and the Kelvin-Planck statements of the Second Law come just as a part of the Second Law, i.e., the encoded part. As we have shown the different results in quantum thermodynamics are obtained just because quantum thermodynamic systems contain quantum correlations through which information can be decoded (brought back) spontaneously without any demon intervening in the process and consequently more work than what is expected can be output. Decoding information (negative entropy production), without a demon, is never seen in deterministic classical thermodynamics therefore the second law of classical thermodynamics cannot be extended to quantum thermodynamics. In the next section we will show that there is a thermodynamic force which is responsible for decoding and encoding information.
\newline
\section{Maxwell's demon and quantum thermodynamic force}
In Ref. \cite{Ahmadi} it was shown that a thermodynamic force is responsible for the flow and backflow of information in quantum processes. For a system, interacting with a bath initially at temperature $\beta=1/T$, the rate of the entropy production can be expressed as \cite{Ahmadi}
\begin{equation}\label{eq:15}
\dfrac{d_iS}{dt}=tr\{F_{th}V_{th}\},
\end{equation}
where $V_{th}\equiv\dot{\rho}_t\rho_t^\beta$ is the thermodynamic flow and $F_{th}\equiv\dfrac{1}{\rho^\beta_t}[\ln\rho^\beta_t-\ln\rho_t]$ the thermodynamic force. Using Eqs. (\ref{eq:a3}) and (\ref{eq:15}) we get
\begin{equation}\label{eq:16}
\dfrac{dW_{irr}}{dt}=\dfrac{1}{\beta} tr\{F_{th}V_{th}\}.
\end{equation}
Since it was shown in Ref. \cite{Ahmadi} that the thermodynamic force $F_{th}$ is responsible for the flow (encoding) and backflow (decoding) of information in Markovian and non-Markovian dynamics, respectively, Eq. (\ref{eq:16}) suggests that, if the system is left to itself, $F_{th}$ actually encodes energy, during the flow, not to be used as work by the system and decodes energy, during the backflow, to be used as work by the system. In classical thermodynamics De Donder found a similar relation for chemical reactions \cite{Donder}. Let us now consider the case in which the system is not left to itself, i.e., someone or something outside the system (as a demon) intervenes in the process. Szil\'{a}rd argued that negative work $\Delta W$ can be extracted from an isothermal cycle if Maxwell's demon plays the role of a feedback controller \cite{Nielsen}. When the statistical state of a system changes from $\rho(x)$ to $\rho(x|m)$, due to the measurements made by the demon on the system, the change in the entropy of the system can be expressed as \cite{Parrondo,Lloyd1}
\begin{equation}\label{eq:16a}
\Delta S_{meas}=H(X|M)-H(X)=-I(X:M),
\end{equation}
where $H(X)=-\sum_x\rho(x)\ln\rho(x)$ is the Shanon entropy of the system and $I(X;M)$ the mutual information between the state of the system and the measurement outcome $M$. Since $I(X;M)$ is always positive thus the demon causes the entropy of the system to decrease. This is similar to the case of non-Markovianity in which the entropy decreases. Therefore the presence of the demon is also expected to lead to extracting more work from the system than what is expected. Now the role of the demon can be incorporated into the Second Law as \cite{Takahiro,Parrondo}
\begin{equation}\label{eq:16b}
\Delta W\geq\Delta F-\dfrac{1}{\beta}I(X:M).
\end{equation}
In Ref. \cite{Sagawa} a practical way was offered, as an alternative to the Szil\'{a}rd engine, to physically realize Maxwell's demon. They have shown that using a feedback contoller (the demon) which makes measurements on the engine they are capable of extracting more work from the heat reservoirs than is otherwise possible in thermal equilibrium. For a system, initially and finally in equilibrium states with temperature $\beta=1/T$, which can contact heat reservoirs $B_1, B_2, ..., B_n$ at respective temperatures $T_1, T_2, ..., T_n$ they have found that
\begin{equation}\label{eq:17}
\Delta W \geq\Delta F^\beta-\dfrac{1}{\beta}I(\rho_1:X),
\end{equation}
and
\begin{equation}\label{eq:17b}\nonumber
I(\rho_1:X)=\dfrac{1}{\beta}[S(\rho_1)-H(\{p_k\})+H(\rho_1:X)],
\end{equation}
where $\rho_1$ is the state of the system at some time $t_1$, $S(\rho_1)$ the Von Neumann entropy, $H(\{p_k\})=-\sum_{k}p_k\ln p_k$ the Shannon information content and $H(\rho_1:X)=-\sum_{k}tr\{\sqrt{D_k}\rho_1\sqrt{D_k}\ln\sqrt{D_k}\rho_1\sqrt{D_k}\}$. $\{D_k\}$ are positive operator valued-measure (POVM) defined by $D_k=M^\dagger_kM_k$ and $p_k=tr\{D_k\rho\}$. It is seen that the sum of the last three terms on the right hand side of the inequality (\ref{eq:17}) is the irreversible work due to the presence of the feedback controller (the demon). Thus if we take the time derivative of these three terms we have
\begin{eqnarray}\label{eq:18}\nonumber
\dfrac{dW^{dem}_{irr}}{dt}&=&\dfrac{1}{\beta}[tr\{\dot{\rho_1}\ln\rho_1\}+\sum_{k}\dot{p}_k\ln p_k\\
&-&\sum_{k}tr\{\sqrt{D_k}\dot{\rho}_1\sqrt{D_k}\ln\sqrt{D_k}\rho_1\sqrt{D_k}\}].
\end{eqnarray}
Comparing Eq. (\ref{eq:18}) with Eq. (\ref{eq:16}) it is observed that there are three quantum thermodynamic forces responsible for the extra work done during the process,
\begin{equation}\label{eq:19}
F^1_{th}=\dfrac{\ln\rho_1}{\rho^\beta_1},\ F^{2(k)}_{th}=\dfrac{\ln p_k}{p^\beta_k},\ F^{3(k)}_{th}=-\dfrac{\ln\sqrt{D_k}\rho_1\sqrt{D_k}}{\rho^\beta_1}.
\end{equation}
Thus we may write
\begin{equation}\label{eq:20}
F^{tot}_{th}=F^1_{th}\bigoplus F^2_{th}\bigoplus F^3_{th}.
\end{equation}
There are also three thermodynamic flows associated with these three thermodynamic forces above,
\begin{equation}\label{eq:21}
V^1_{th}=\dot{\rho_1}\rho^\beta_1,\ V^{2(k)}_{th}=\dot{p}_kp^\beta_k,\ V^{3(k)}_{th}=\sqrt{D_k}\dot{\rho}_1\sqrt{D_k}\rho^\beta_1,
\end{equation}
and it may be written
\begin{equation}\label{eq:22}
V^{tot}_{th}=V^1_{th}\bigoplus V^2_{th}\bigoplus V^3_{th}.
\end{equation}
We must notice that Eqs. (\ref{eq:20}) and (\ref{eq:22}) should not be taken too literally, i.e., these equations just indicate the fact that there are three thermodynamic forces and flows involved due to the presence of the feedback controller and we cannot add them up like the way we do about typical vectors. We note that $F^{tot}_{th}=0$ if and only if $D_k$ is proportional to the identity operator for all $k$ \cite{Sagawa}, which means that nothing is intervening in the process, therefore no information is decoded to be used to perform additional work by the system. On the other hand, $F^{tot}_{th}=F^2_{th}$ if and only if $D_k$ is the projection operator satisfying $[\rho_1, D_k]=0$ for all $k$ \cite{Sagawa}, which means that the measurement on $\rho_1$ is classical, hence $F^{tot}_{th}$ is classical. In Refs. \cite{Parrondo,Jung,Ren} similar results have been found. Therefore we have shown that intervention (the demon) from the outside in the process of a system may be represented by a thermodynamic force.
\section{Summary}
In this work we have appropriately divided the work done by a thermodynamic system into two parts: reversible work and irreversible work. This partitioning seems plausible since whenever the process is reversible all the work is reversible and there exists no irreversible work as expected. Using this partitioning we have derived a generic form for the efficiency of an engine operating in an arbitrary cycle. It was shown that negative entropy production, which can occur in non-Markovian processes or by intervening in the process of the system (Maxwell's demon), means that the internal energy is decoded to be used by the system to perform more work than what is expected and this additional work leads to having quantum engines with efficiencies greater than that of Carnot. We have investigated four special cases to elucidate the physical meaning of $\Delta W_{irr}=\dfrac{1}{\beta}\Delta_iS$ in quantum thermodynamics and have discovered results which strictly contradict the second law of classical thermodynamics. We have also shown that the relation $\Delta W_{irr}=\dfrac{1}{\beta}\Delta_iS$ as a link between thermodynamics and information theory is fundamental. Based on this analysis we have introduced a new definition of the second law of thermodynamics such that it covers both classical and quantum thermodynamics and incorporates well information into the Second Law. At last, we have shown that a quantum thermodynamic force is responsible for encoding and decoding information even when a feedback controller outside the system is involved in the process.
\section*{ACKNOWLEDGMENTS}
The authors would like to thank Profs. Johan {\AA}berg and Michal Horodecki for very useful comments and advices.
%========================================================================
%========================================================================
%========================================================================

\section{Supplementary Note 1}\label{Note1}
The entropy production between the engine and the hot reservoir is
\begin{equation}\label{n1}
\Delta_iS_1=\dfrac{\Delta Q_h}{T_1}-\dfrac{\Delta Q_h}{T_h},
\end{equation}
and between the engine and the cold reservoir
\begin{equation}\label{n2}
\Delta_iS_2=\dfrac{\Delta Q_c}{T_c}-\dfrac{\Delta Q_c}{T_3}.
\end{equation}
The total entropy production during a cycle is $\Delta_iS=\Delta_iS_1+\Delta_iS_2$. Since no irreversibility occurs in the interior of the engine we have
\begin{equation}\label{n3}
\dfrac{\Delta Q_h}{T_1}=\dfrac{\Delta Q_c}{T_3}.
\end{equation}
Thus we get
\begin{equation}\label{n4}
\Delta_iS=\dfrac{\Delta Q_c}{T_c}-\dfrac{\Delta Q_h}{T_h}.
\end{equation}
The efficiency of the engine reads
\begin{equation}\label{n5}
\eta_e\equiv\dfrac{-(\Delta W_1+\Delta W_2)}{\Delta Q_h}=1-\dfrac{\Delta Q_c}{\Delta Q_h}=1-\dfrac{T_3}{T_1}.
\end{equation}
Now combining Eqs. (\ref{n4}) and (\ref{n5}) we obtain
\begin{equation}\label{n6}
\eta_e-\eta_C=-\dfrac{T_c\Delta_iS}{\Delta Q_h},
\end{equation}
where $\eta_C=1-\dfrac{T_c}{T_h}$ is the Carnot efficiency.
\section{Supplementary Note 2}\label{Note2}
Irreversibility in physical process is strictly related to the idea of energy dissipation. Irreversible processes encountered by an open thermodynamic system are accompanied with a production of entropy which is fundamentally different from the entropy flow in the form of heat caused by the interaction between the system and its environment. Characterization of irreversibility is one of the cornerstones of non-equilibrium thermodynamics since the theory was born. For an isothermal process the second law of deterministic equilibrium classical thermodynamics may be expressed as
\begin{equation}\label{eq:1}
\Delta W\geq \Delta F^\beta,
\end{equation}
where $\Delta W$ is the amount of work required to change the state of the system between two equilibrium states and $\Delta F^\beta$ the difference in the Helmholtz free energy of the system. This, in turn, led to the introduction of the so-called irreversible work \cite{Jarzynski,Crooks},
\begin{equation}\label{eq:2}
\Delta W_{irr}\equiv\Delta W-\Delta F^\beta\geq0.
\end{equation}
Defining $\Delta W_{irr}$ as in Eq. (\ref{eq:2}) gives rise to
\begin{equation}\label{eq:3}
\Delta W_{irr}=\dfrac{1}{\beta}\Delta_iS,
\end{equation}
where $\Delta_iS$ is the entropy production of the system during the irreversible process and $\beta=1/T$ the temperature of the system. It should be emphasised that in equilibrium thermodynamics the reversible work equals the change in the free energy, i.e, $\Delta W_{rev}=\Delta F^\beta$ \cite{Kondepudi,Blundell,Callen}. Thus the total work done by the system is partitioned into reversible and irreversible parts, i.e, $\Delta W=\Delta W_{rev}+\Delta W_{irr}$. Thermodynamic reversibility is achieved if and only if no entropy is produced inside the system, i.e, $\Delta_iS=0$ \cite{Kondepudi,Blundell,Callen}. Eq. (\ref{eq:3}), in the language of information, has a very subtle and interesting meaning. It links thermodynamics with information theory. It says that some of the internal energy, during the irreversible process, is encoded due to the loss of information and consequently the system cannot use this amount of the internal energy to do work. For instance in the process of the free expansion of a gas all the internal energy is encoded therefore no internal energy can be used by the gas to perform any work, i.e, $|\Delta W_{rev}|=|\Delta W_{irr}|$. Thus if the system operates in an irreversible cycle its efficiency decreases \cite{Kondepudi,Blundell,Callen}. In other words Eq. (\ref{eq:3}) means that information is physical. Eqs. (\ref{eq:2}) and (\ref{eq:3}) have been extended to quantum thermodynamics with the same formulae \cite{Broeck,Deffner,Esposito,Plastina,Souza,Francica,Campbell,Tiago,Lutz}. This may be wrong and misleading for two reasons. First, the inequality (\ref{eq:1}) does not always hold in the quantum realm. Because in stochastic quantum thermodynamics there exist processes, called non-Markovian processes, in which we may have
\begin{equation}\label{eq:3a}\nonumber
\Delta W\leq\Delta F^\beta.
\end{equation}
This is because in deterministic classical thermodynamics, according to the Clausius' statement of the Second Law, the entropy production of a system can never be negative \cite{Kondepudi,Blundell,Callen}, but in quantum thermodynamics, during non-Markovian processes, the entropy production of the system may be negative \cite{Marcantoni,Chen}. Second, in quantum thermodynamics, as we will show in this work, the reversible work $\Delta W_{rev}$ done by a system equals $\dfrac{1}{\beta}\Delta I+\Delta F^\beta$, where $I(t)=S(\rho_t\|\rho^\beta_t)$, rather than $\Delta F^\beta$. If we apply Eq. (\ref{eq:2}) to the evolution of a closed quantum system, which is unitary, we observe that $\Delta W_{irr}$ could be nonzero while the process is reversible, i.e, $\Delta_iS=0$. Thus if $\Delta W_{irr}$ is defined as in Eq. (\ref{eq:2}) for quantum systems, for a closed quantum system Eq. (\ref{eq:3}) may not hold. For instance in the case of the evolution of a closed quantum system initially in equilibrium we have $\Delta W_{irr}=\dfrac{1}{\beta}S(\rho_t\|\rho^\beta_t)\neq0$ while $\Delta_iS=0$.
%Or in the case of thermal operations we have $\Delta W_{irr}=0$, since $\Delta W=\Delta F^\beta=0$, while the entropy is produced during the process, i.e, $\Delta_iS\neq0$ \cite{Santos}. Eq. (\ref{eq:2}) in stochastic thermodynamics is the energetic deviation from an equilibrium process and is better to be called the non-equilibrium work, i.e., $\Delta W_{neq}$.
\newline
\textbf{Reversible and irreversible work}. We define heat as
\begin{equation}\label{eq:4}
\Delta Q\equiv\int_{0}^{\tau}dttr\{\dot{\rho}_tH_t\},
\end{equation}
and work is defined as the mean change of Hamilton with time \cite{Alicki}
\begin{equation}\label{eq:5}
\Delta W\equiv\int_{0}^{\tau}dttr\{\rho_t\dot{H}_t\},
\end{equation}
where $\rho_t$ is the state of the system and $H_t$ the Hamiltonian of the system. Now consider an arbitrary quantum system $S$ coupled with a heat reservoir $B$ at temperature $\beta=1/T$. Eq. (\ref{eq:5}) becomes
\begin{equation}\label{eq:6}
\Delta W=-\dfrac{1}{\beta}\int_{0}^{\tau}dttr\{\rho_t\partial_t\ln\rho^{\beta}_t\}+\Delta F^{\beta},
\end{equation}
where $\rho^{\beta}_t=\exp(-\beta H_t)/Z_t$ is the instantaneous Gibbs state of the system with $Z_t$ the partition function and $F^{\beta}_t=-\dfrac{1}{\beta}\ln Z_t$ the free energy of the system. The total change in the entropy $\Delta S$ of the system is divided into two parts \cite{Kondepudi,Blundell,Callen}
\begin{equation}\label{eq:7}
\Delta S=\Delta_iS+\Delta_eS,
\end{equation}
in which $S=-tr\{\rho\ln\rho\}$ is the Von Neumann entropy of the system, $\Delta_eS\equiv \beta \Delta Q$ the entropy change due to the exchange of energy with the reservoir and $\Delta_iS$ the entropy produced by the irreversible processes in the interior of the system. In contrast to the thermodynamic entropy that can be defined only for thermal equilibrium, the Von Neuwman entropy can be defined for an arbitrary probability distribution. Combining Eqs. (\ref{eq:4})$-$(\ref{eq:7}), we get
\begin{equation}\label{eq:8}
\Delta_iS=S(\rho_0\|\rho_0^{\beta})-S(\rho_\tau\|\rho_\tau^{\beta})-\int_{0}^{\tau}dttr\{\rho_t\partial_{t}\ln\rho^{\beta}_t\},
\end{equation}
where $S(\rho\|\sigma)\equiv tr\{\rho\ln\rho\}-tr\{\rho\ln\sigma\}$ is the relative entropy of the states $\rho$ and $\sigma$. A thermodynamic reversible process is defined as a process that can be reversed without leaving any trace on the surroundings. That is, both the system and the surroundings are returned to their initial states at the end of the reverse process. This definition of reversibility in conventional thermodynamics may be completely characterized by the entropy production. Thermodynamic reversibility is achieved if and only if the entropy production is zero, i.e., $\Delta_iS=0$ \cite{Kondepudi,Blundell,Callen}. A stochastic process is thermodynamically reversible, if and only if the final probability distribution can be restored to the initial one, without remaining any effect on the outside world \cite{Takahiro}. As in conventional thermodynamics, reversibility in stochastic processes is completely characterized by the entropy production. Reversibility in stochastic thermodynamics is achieved if and only if the entropy production is zero \cite{Takahiro}, i.e.,
\begin{equation}\label{eq:8a}
\Delta_iS=0.
\end{equation}
\begin{thm}\label{Theorem}
The work done by a thermodynamic system, in the weak coupling limit, can always be appropriately partitioned into two parts: reversible work and irreversible work, i.e,
\begin{equation}\label{eq:9}
\Delta W=\Delta W_{rev}+\Delta W_{irr},
\end{equation}
in which
\begin{equation}\label{eq:10}
\Delta W_{rev}=\dfrac{1}{\beta}\Delta I+\Delta F^\beta,
\end{equation}
and
\begin{equation}\label{eq:11}
\Delta W_{irr}=\dfrac{1}{\beta}\Delta_iS.
\end{equation}
\end{thm}
\textbf{Proof.} Since in a reversible process no entropy is produced inside the system, i.e., $\Delta_iS=0$, using Eqs. (\ref{eq:6}) and (\ref{eq:8}), after some straightforward calculations, the (reversible) work is obtained as
\begin{equation}\nonumber
\Delta W_{rev}=\dfrac{1}{\beta}\Delta I+\Delta F^\beta,
\end{equation}
where $I(t)=S(\rho_t\|\rho^\beta_t)$. Unlike the reversible processes, during a general process the entropy may be produced inside the system (irreversible processes), i.e., $\Delta_iS\neq0$ \cite{Gemmer,Breuer}. Therefore we find
\begin{equation}\nonumber
\Delta W=\dfrac{1}{\beta}\Delta_iS+\dfrac{1}{\beta}\Delta I+\Delta F^\beta,
\end{equation}
where the sum of the last two terms on the right hand side is the reversible work, $\Delta W_{rev}$, and the first term is the irreversible work, $\Delta W_{irr}$. Hence the total work done by a system during a general process can be expressed as
\begin{equation}\nonumber
\Delta W=\Delta W_{irr}+\Delta W_{rev}.\ \ \ \ \ \  \Box
\end{equation}
The non-equilibrium free energy for a generic statistical state $\rho$ of a quantum system in contact with a thermal bath is defined as
\begin{equation}\label{eq:11a}
F(\rho, H)\equiv E-TS=tr\{\rho H\}-TS(\rho),
\end{equation}
where $H$ is the Hamilton of the system. Using Eqs. (\ref{eq:4}) and (\ref{eq:5}), Theorem \ref{Theorem}, and Eq. (\ref{eq:11a}) we obtain
\begin{equation}\label{eq:11b}
\dfrac{1}{\beta}\Delta_iS=\Delta W_{irr}=\Delta W-\Delta F.
\end{equation}
Eq. (\ref{eq:11b}) is the extension of Eq. (\ref{eq:2}) to quantum thermodynamics. The only difference is that $\Delta F$ in Eq. (\ref{eq:11b}) is the difference in non-equilibrium free energies and as we mentioned before this is because quantum thermodynamics is a non-equilibrium thermodynamics. The associated non-equilibrium free energy is analogous to its equilibrium counterpart in non-equilibrium processes. When the initial and final states of the system are thermal equilibrium states Eq. (\ref{eq:11b}) becomes equivalent to Eq. (\ref{eq:2}) in conventional thermodynamics as expected.
\section{Supplementary Note 3}\label{Note3}
Consider a system in state $\rho_0$ at time $t=0$ attached to a bath of temperature $T$. After a finite-time $\tau$, let the state of the system be $\rho_\tau$. The Hamiltonian $H$ of the system remains unchanged during the evolution. Therefore, using Eq. (\ref{eq:8}), the entropy production of the system after a time $\tau$ is
\begin{equation}\label{eq:11d}
\Delta_iS=S(\rho_0\|\rho^\beta)-S(\rho_\tau\|\rho^\beta).
\end{equation}
For a completely positive, trace preserving (CPTP) map $\Lambda_t$ and any two density matrices $\rho_1$ and $\rho_2$, if the dynamics is Markovian for which $\Lambda_t[\rho^\beta]=\rho^\beta$ for all $t$, we have \cite{Rivas}
\begin{equation}\label{eq:11e}
S(\rho_2|\rho_\beta)=S(\Lambda_t[\rho_1]\|\Lambda_t[\rho^\beta])\leq S(\rho_1\|\rho^\beta).
\end{equation}
But if the dynamics is non-Markovian, since $\Lambda_t[\rho^\beta]\neq\rho^\beta$, we can have \cite{Rivas}
\begin{equation}\label{eq:11f}
S(\rho_2\|\rho^\beta)\geq S(\rho_1\|\rho_\beta).
\end{equation}
The heat exchanged between the system and the bath is obtained as
\begin{eqnarray}\label{eq:11g}\nonumber
\Delta Q&=&T\Delta S-T\Delta_iS \nonumber \\
&=&T[S(\rho_\tau)-S(\rho_0)]+T[S(\rho_\tau\|\rho^\beta)-S(\rho_0\|\rho^\beta)].\nonumber \\
\end{eqnarray}
\section{Supplementary Note 4}\label{Note4}
Here we consider a spin-1/2 system \cite{Quan,Thomas,Thomas1,Kieu} working in an Otto cycle, as depicted in Fig. (\ref{fig:Fig2.pdf}). The system is in an initial state $\rho_0$, diagonal in the eigenbasis of the Hamiltonian $H_0=(\omega_0/2)\sigma_z$, where $\omega_0=\kappa B$ and $\sigma_z$ is the Pauli matrix. Here $\kappa$ is a constant and $B$ is the constant magnetic field applied in the $z$ direction on the system. In step I the engine interacts weakly with a hot reservoir at temperature $T_h$, for time $\tau_1$ from point $A(\rho_0, H_0)$ to point $B(\rho_1, H_0)$. The final state of the system is $\rho_1=\exp(-H_0/T_1)/tr\{\exp(-H_0/T_1)\}$, which is diagonal in the eigenbasis of $H_1$. Here $T_1=-\omega_0/(2\tanh^{-1}\langle\sigma_z\rangle_1)$, where $\langle\sigma_z\rangle=tr\{\rho_1\sigma_z\}$, is the effective temperature of the system after time $\tau_1$. The heat absorbed by the engine is $\Delta Q_h=tr\{H_0(\rho_1-\rho_0)\}$. In step II the engine is decoupled from the hot reservoir and undergoes an adiabatic evolution from point $B(\rho_1, H_0)$ to point $C(\rho_1, H_1)$ by varying the magnetic field from $\omega_0$ to $\omega_1$ ($\omega_1<\omega_0$). Since the system performs work, the temperature of the system changes at the end of this process and it becomes $T_2=T_1\omega_1/\omega_0$. In step III it interacts weakly with a cold reservoir at temperature $T_c$ from point $C(\rho_1, H_1)$ to point $D(\rho_0, H_1)$ for time $\tau_2$ and the state of the system becomes $\rho_0$ with the effective temperature $T_3=-\omega_1/(2\tanh^{-1}\langle\sigma_z\rangle_0)$. The heat rejected to the cold reservoir is $\Delta Q_c=tr\{H_1(\rho_0-\rho_1)\}$. Finally in step IV the engine is decoupled from the cold reservoir and, in an adiabatic evolution, goes back to its initial point by going from point $D(\rho_0, H_1)$ to point $A(\rho_0, H_0)$ and complete the cycle. The temperature of the system at the end of this cycle becomes $T_0=T_3\omega_0/\omega_1$. It can be shown that the effective temperatures of the system approach the temperatures of the heat baths asymptotically \cite{Thomas}. The heat absorbed by the system from the hot reservoir during step I is given by
\begin{equation}\label{14aa}
\Delta Q_h=\dfrac{\omega_0}{2}[\tanh(\dfrac{\omega_1}{2T_3})-\tanh(\dfrac{\omega_0}{2T_1})].
\end{equation}
In the same way, the heat rejected to the cold heat during step III is obtained as
\begin{equation}\label{14aa1}
\Delta Q_c=-\dfrac{\omega_1}{2}[\tanh(\dfrac{\omega_1}{2T_3})-\tanh(\dfrac{\omega_0}{2T_1})].
\end{equation}
Now the total work done by the system after the cycle is $\Delta W=-(\Delta Q_h+\Delta Q_c)$, i.e,
\begin{equation}\label{14aa2}
\Delta W=\dfrac{\omega_1-\omega_0}{2}[\tanh(\dfrac{\omega_1}{2T_3})-\tanh(\dfrac{\omega_0}{2T_1})].
\end{equation}
For a machine to work as an engine we must have $\Delta W<0$, $\Delta Q_h>0$ and $\Delta Q_c<0$. This implies that
\begin{equation}\label{14aa3}
\dfrac{\omega_1}{T_3}\geq\dfrac{\omega_0}{T_1}.
\end{equation}
Hence the efficiency of the engine reads
\begin{equation}\label{14a}
\eta=1-\dfrac{\omega_1}{\omega_0}\leq1-\dfrac{T_3}{T_1}.
\end{equation}

\begin{references}
\bibitem{Kondepudi} Kondepudi D and Prigogine I, Modern Thermodynamics (New York: Wiley 1998).
\bibitem{Blundell} S. J. Blundell, K. M. Blundell, Concepts in Thermal Physics (Oxford University Press 2009).
\bibitem{Callen} H. B. Callen, Thermodynamics and an Introduction to Thermostatistics 2nd edn
(JohnWiley, 1985).
\bibitem{Maxwell} J. C. Maxwell, Theory of Heat (Appleton, London, 1871).
\bibitem{Leff} H. S. Leff and A. F. Rex, (eds) in Maxwell's Demon: Entropy, Information,
Computing (Princeton Univ. Press, 1990).
\bibitem{Maruyama} K. Maruyama, F. Nori, and V. Vedral, \textcolor{blue}{Rev. Mod. Phys. \textbf{81}, 1-23 (2009)}.
\bibitem{Evans}D. J. Evans, D. Cohen, and G. P. Morriss, \textcolor{blue}{Phys. Rev. Lett. \textbf{71}, 2401–2404 (1993)}.
\bibitem{Searles}D. J. Evans, and D. J. Searles, \textcolor{blue}{Phys. Rev. E \textbf{40}, 1645–1648 (1994)}.
\bibitem{Wang}G. M. Wang, E. M. Sevick, E. Mittag, D. J. Searles, and D. J. Evans, \textcolor{blue}{Phys. Rev. Lett. \textbf{89}, 050601 (2002)}.
\bibitem{An}S. An, J.-N. Zhang, M. Um, D. Lv, Y. Lu, J. Zhang, Z.-Q. Yin, H. T. Quan, and K. Kim, \textcolor{blue}{Nature Phys. \textbf{11}, 193–199 (2015)}.
\bibitem{Zeitschrift} L. Szil\'{a}rd, \textcolor{blue}{Zeitschrift f\"{u}r Physik \textbf{53}, 840 (1929)}.
\bibitem{Chen} Hong-Bin Chen, Guang-Yin Chen, and Yueh-Nan Chen, \textcolor{blue}{Phys. Rev. A \textbf{96}, 062114 (2017)}.
\bibitem{Ahmadi} B. Ahmadi, S. Salimi, A. S. Khorashad and F. Kheirandish, \textcolor{blue}{Sci. Rep. \textbf{9}, 8746 (2019)}.
\bibitem{Quan} Quan, H. T., Liu, Y.-x., Sun, C. P. and Nori, F., \textcolor{blue}{Phys. Rev. E \textbf{76}, 031105 (2007)}.
\bibitem{Marcantoni} S. Marcantoni, S. Alipour, F. Benatti, R. Floreanini, and A. T. Rezakhani, \textcolor{blue}{Sci. Rep. \textbf{7}, 12447 (2017)}.
\bibitem{Gemmer} J. Gemmer, M. Michel and G. Mahler, Quantum Thermodynamics, Lect. Notes Phys. 784 (Springer, Berlin Heidelberg 2009)
\bibitem{Thomas} G. Thomas, N. Siddharth, S. Banerjee, and S. Ghosh, \textcolor{blue}{Phys. Rev. E \textbf{97}, 062108 (2018)}.
\bibitem{Thomas1} G. Thomas and R. S. Johal, \textcolor{blue}{Phys. Rev. E \textbf{83}, 031135 (2011)}.
\bibitem{Kieu} T. D. Kieu, \textcolor{blue}{Phys. Rev. Lett. \textbf{93}, 140403 (2004)}.
\bibitem{Xiong} H. N. Xiong, W. M. Zhang, X. G. Wang and M. H Wu, \textcolor{blue}{Phys. Rev. A \textbf{82} 012105 90 (2010)}.
\bibitem{Zhang} X. Y. Zhang, X. L. Huang, and X. X. Yi, \textcolor{blue}{J. Phys. A: Math. Theor. \textbf{47}, 455002 (2014)}.
\bibitem{Paavola} J. Paavola, J. Piilo, K.-A. Suominen, and S. Maniscalco, \textcolor{blue}{Phys. Rev. A \textbf{79}, 052120 (2009)}.
\bibitem{Intravaia} F. Intravaia, S. Maniscalco, and A. Messina, \textcolor{blue}{Phys. Rev. A \textbf{67}, 042108 (2003)}.
\bibitem{Goan}  H. S. Goan, P. W. Chen, and C. C. Jian, \textcolor{blue}{J. Chem. Phys. \textbf{134}, 124112 (2011)}.
\bibitem{Deffner1} S. Deffner and C. Jarzynski, \textcolor{blue}{Phys. Rev. X \textbf{3}, 041003 (2013)}.
\bibitem{Landauer} R. Landauer, Information is physical, \textcolor{blue}{Physics Today, 44(5), 23-29 (1991)}.
\bibitem{Karnani} M. Karnani, K. Pääkkönen, and A. Annila, \textcolor{blue}{Pro. R. Soc. A \textbf{465} (2107), 2155-2175 (2009)}.
\bibitem{Vedral} V. Vedral, Decoding reality: The universe as quantum information, Oxford University Press (2010).
\bibitem{Vedral1} V. Vedral,  Information and physics. \textcolor{blue}{Information, 3(2), 219-223 (2012)}.
\bibitem{Davies} P. Davies, and N. H. Gregersen, Information and the nature of reality: from physics to metaphysics, Cambridge University Press (2010).
\bibitem{Rio} del Rio, L., {\AA}berg, J., Renner, R., Dahlsten, O. and Vedral, \textcolor{blue}{Nature \textbf{474}, 61-63 (2011)}.
\bibitem{Sagawa} T. Sagawa and M. Ueda, \textcolor{blue}{Phys. Rev. Lett. \textbf{100}, 080403 (2008)}.
\bibitem{Landauer1} R. Landauer, \textcolor{blue}{IBM J. Res. Dev. \textbf{5}, 183 (1961)}.
\bibitem{Parrondo} Juan M. R. Parrondo, Jordan M. Horowitz and Takahiro Sagawa, \textcolor{blue}{Nature Phys. \textbf{11}, 131-139 (2015)}.
\bibitem{Lloyd} S. Lloyd, \textcolor{blue}{Phys. Rev. A \textbf{56}, 3374 (1997)}.
\bibitem{Donder} De Donder, T., Van Rysselberghe, P., Affinity. 1936, Stanford University Press: Menlo Park, CA.
\bibitem{Nielsen} M. A. Nielsen, C. M. Caves, B. Schumacher, and H. Barnum, \textcolor{blue}{Proc. R. Soc. A \textbf{454}, 277 (1998)}.
\bibitem{Lloyd1} S. Lloyd \textcolor{blue}{Phys. Rev. A \textbf{39}, 5378-5386 (1989)}.
\bibitem{Takahiro} Takahiro Sagawa, \textcolor{blue}{arXiv:1712.06858}.
\bibitem{Jung} Jung Jun Park, Kang-Hwan Kim, T. Sagawa, and Sang Wook Kim, \textcolor{blue}{Phys. Rev. Lett. \textbf{111}, 230402 (2013)}.
\bibitem{Ren} Li-Hang Ren and Heng Fan, \textcolor{blue}{Phys. Rev. A \textbf{96}, 042304 (2017)}.
\bibitem{Jarzynski} C. Jarzynski, \textcolor{blue}{Phys. Rev. Lett. \textbf{78}, 2690 (1997)}.
\bibitem{Crooks} G. E. Crooks, \textcolor{blue}{Phys. Rev. E \textbf{60}, 2721 (1999)}.
\bibitem{Broeck} J. M. R. Parrondo, C. Van den Broeck, and R. Kawai, \textcolor{blue}{New J. Phys. \textbf{11}, 073008 (2009)}.
\bibitem{Deffner} S. Deffner and E. Lutz, \textcolor{blue}{Phys. Rev. Lett. \textbf{105}, 170402 (2010)}.
\bibitem{Esposito} M. Esposito and C. Van den Broeck,  \textcolor{blue}{EPL, \textbf{95} (2011) 40004}.
\bibitem{Plastina} F. Plastina, A. Alecce, T. J. G. Apollaro, G. Falcone, G. Francica, F. Galve, N. Lo Gullo, and R. Zambrini,  \textcolor{blue}{Phys. Rev. Lett. \textbf{113}, 260601 (2014)}.
\bibitem{Souza} T. B. Batalh\~{a}o, A. M. Souza, R. S. Sarthour, I. S. Oliveira, M. Paternostro, E. Lutz, and R. M. Serra, \textcolor{blue}{Phys. Rev. Lett. \textbf{115}, 190601 (2015)}.
\bibitem{Francica} G. Francica, J. Goold, and F. Plastina,  \textcolor{blue}{arXiv:1707.06950}.
\bibitem{Campbell} S. Deffner and S. Campbell, \textcolor{blue}{J. Phys. A: Math. Theor (2017)}.
\bibitem{Tiago} T. B. Batalh\'{a}o, S. Gherardini, J. P. Santos, G. T. Landi, and M. Paternostro,  \textcolor{blue}{arXiv:1806.08441}.
\bibitem{Lutz} S. Deffner and E. Lutz, \textcolor{blue}{Phys. Rev. Lett. \textbf{107}, 140404 (2011)}.
%\bibitem{Santos} J. P. Santos, L. C. C´eleri, G. T. Landi, and M. Paternostro,  \textcolor{blue}{npj Quant. Inf. \textbf{5}, 23 (2019)}.
\bibitem{Alicki} R. Alicki, \textcolor{blue}{J. Phys. A \textbf{12}, L103 (1979)}.
\bibitem{Breuer} H. P. Breuer and F. Petruccione, The theory of open quantum systems (Oxford University Press, Oxford, 2002).
\bibitem{Rivas} A. Rivas, S. F. Huelga and M. B. Plenio, \textcolor{blue}{Rep. Prog. Phys. \textbf{77}, 094001 (2014)}.
\end{references}
\end{document}